\UseRawInputEncoding
\documentclass[lettersize,journal]{IEEEtran}
\pdfoutput=1
\usepackage{amsmath,amsfonts}
\usepackage{algorithmic}
\usepackage{algorithm}
\usepackage{array}
\usepackage[caption=false,font=normalsize,labelfont=sf,textfont=sf]{subfig}
\usepackage{textcomp}
\usepackage{stfloats}
\usepackage{url}
\usepackage{verbatim}
\usepackage{graphicx}
\usepackage{cite}
\usepackage{bm}
\usepackage[justification=centering]{caption}

\begin{document}

\title{A Fast Alternating Minimization Algorithm for Coded Aperture Snapshot Spectral Imaging Based on Sparsity and Deep Image Priors}

\author{Qile Zhao, Xianhong Zhao, Xu Ma, \IEEEmembership{Senior Member, IEEE}, Xudong Chen, \IEEEmembership{Fellow, IEEE}, Gonzalo R. Arce, \IEEEmembership{Fellow, IEEE}
\thanks{Q. Zhao, X Zhao and X. Ma are with the School of Optics and Photonics, Key Laboratory of Photoelectronic Imaging Technology and System of Ministry of Education of China, Beijing Institute of Technology, Beijing 100081, China (e-mail: qile.zhao@outlook.com; garyzhao97@163.com; maxu@bit.edu.cn). (\emph{Corresponding author: Xu Ma}). }
\thanks{X. Chen is with the Department of Electrical and Computer Engineering, National University of Singapore, Singapore 117583 (e-mail: elechenx@nus.edu.sg).}
\thanks{G. R. Arce is with the Department of Electrical and Computer Engineering, University of Delaware, Newark, DE 19716, USA (e-mail:  arce@udel.edu).}
}

\markboth{This work has been submitted to the IEEE for possible publication. Copyright may be transferred without notice. }%
{Shell \MakeLowercase{\textit{et al.}}: A Sample Article Using IEEEtran.cls for IEEE Journals}


\maketitle

\begin{abstract}
	Coded aperture snapshot spectral imaging (CASSI) is a technique used to reconstruct three-dimensional hyperspectral images (HSIs) from one or several two-dimensional projection measurements. However, fewer projection measurements or more spectral channels leads to a severly ill-posed problem, in which case regularization methods have to be applied. In order to significantly improve the accuracy of reconstruction, this paper proposes a fast alternating minimization algorithm based on the sparsity and deep image priors (Fama-SDIP) of natural images. By integrating deep image prior (DIP) into the principle of compressive sensing (CS) reconstruction, the proposed algorithm can  achieve state-of-the-art results without any training dataset. Extensive experiments show that Fama-SDIP method significantly outperforms prevailing leading methods on simulation and real HSI datasets.
\end{abstract}

\begin{IEEEkeywords}
	hyperspectral imaging, spatial-spectral coding, deep image prior, sparsity prior, compressive sensing, computational imaging.
\end{IEEEkeywords}

\section{Introduction}
\IEEEPARstart{N}{atural} scenes contain rich spectral information, so they can be collected as spatial-spectral three dimensional (3D) cubes named hyperspectral images (HSIs), where two dimensions (2D) represent the spatial domain and another dimension represents the spectral domain. Since HSIs have more information than red-–green–-blue (RGB) images, they have been extensively used in food surveillance\cite{app1}, face recognition\cite{app2}, remote sensing\cite{app3}, biomedical imaging\cite{app4}, etc.
To collect HSIs, conventional imaging approaches use spectrometers to scan scenes along the spatial or spectral dimension, which imposes challenges on the scanning and storage of the datacube. However, the HSIs are highly redundant among the spectral dimension. These challenges may be solved by using compressed sensing (CS)\cite{CSo}. Based on the principles of CS, coded aperture snapshot spectral imaging (CASSI) was proposed\cite{cassi1, cassi2, cassi3, cassi4}. CASSI systems acquire 2D compressive multiplexed projection measurements instead of scanning all voxels in the HSIs. The remarkable advantage of CASSI is that the entire HSIs can be reconstructed with few measurements or even one snapshot. The coded aperture can be optimized to improve the reconstruction performance of CASSI\cite{ca1, ca2}. However, the image reconstruction is an ill-posed problem which  exacerbates as the number of measurements decreases or the spectral channel of HSIs increases. In order to reconstruct 3D HSI from 2D measurements, regularization methods have to be applied\cite{reg}, which typically exploit some prior information of the scenes. In general, effective image priors are critical for CASSI reconstruction, such as total variation (TV)\cite{tv1, tv2}, sparsity\cite{sparse1, sparse2}, low-rank\cite{lowrank1, lowrank2}, and deep image prior (DIP)\cite{dip1, dip2}. In particular, the sparsity prior is not only one of the main principles of CS\cite{cassi2}, but also one of the most salient features of natural images\cite{natural1,natural2}. TV, low-rank and DIP priors are often used for image denoising. DIP has been widely used in recent years due to its excellent denoising ability. 

Several image reconstruction algorithms have been proposed for CASSI. Traditional methods use iterative algorithms based on a regularization term. GPSR\cite{GPSR} and TwIST\cite{TwIST} are used to solve CASSI reconstruction problem based on sparsity prior. GAP-TV\cite{GAP-TV} solves the reconstruction problem with the TV prior. DeSCI\cite{DeSCI} solves the reconstruction problem based on the low-rank property and non-local self-similarity.  However, these traditional algorithms are limited in reconstruction time and performance\cite{DGSMP}. The emergence of deep learning (DL) led to significant improvement in the reconstruction performance of CASSI.
At present, the reconstruction methods based on DL are mainly divided into supervised learning methods and unsupervised learning methods.

\begin{figure*}[!t]
	\centering\includegraphics[width=16cm]{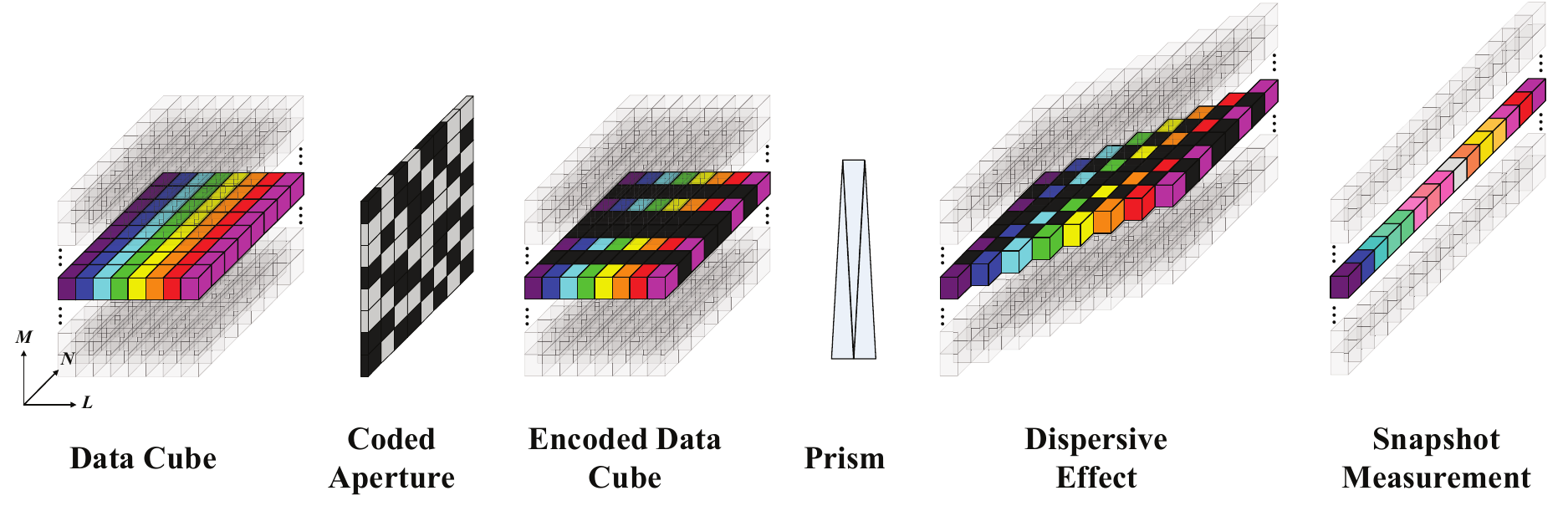}\caption{Schematic of the CASSI system.}
	\label{f1}
\end{figure*}

Supervised leaning methods consider the HSI reconstruction as a non-linear inverse mapping from the compressive measurements to a 3D datacube\cite{rl1,rl2,l-net,TSA-net,PnP,DGSMP}.
$ \lambda $-net\cite{l-net} introduces a two-stage reconstruction network to recover the HSI from a compressive measurement, where the HSI is reconstructed by a self-attention Generative Adversarial Network framework followed by a refinement stage.
TSA-Net\cite{TSA-net} uses three Spatial-Spectral
Self-Attention modules to jointly model the spatial and spectral correlation. PnP-HSI\cite{PnP} firstly trains a denoising network using the hyperspectral dataset, and then performs denoising processing on the results of the GAP-TV algorithm.  DGSMP\cite{DGSMP} introduces an interpretable HSI reconstruction method based on Gaussian scale mixture prior.
The main difference between these supervised-based methods is their network architecture. These approaches treat the inverse imaging as a regression problem, however, the approaches suffers from unsatisfactory reconstruction performance when the number of snapshot is low.
In addition, supervised-based methods are highly dependent on the dataset used. Since hyperspectral data is difficult to collect in large quantities, it is important to develop algorithms based on unsupervised learning. Furthermore, since there may be some small-variations in the acquisition of compressive measurements under different scenes, such as changes in coded aperture or scene noise, the model may not work leading to time-consuming retraining. Therefore, supervised-based methods are often faced with the problem of insufficient generalization ability. 

Due to the aforementioned limitations of CASSI reconstruction algorithms based on supervised learning, the development of unsupervised learning algorithms is important.
PnP-DIP\cite{PNP-DIP} was proposed by the joint use of TV regulation and DIP.
Although PnP-DIP is an iterative algorithm, its final reconstruction result is the output of the DIP network, but the final iterative result of the algorithm is unsatisfactory. 
In addition, since both of the TV and DIP priors used in PnP-DIP are essentially for denoising, the algorithm has limitations in solving ill-posed inverse problems and can easily fall into a local minimum. In order to solve this problem, it is necessary to re-initialize the network parameters and increase the iteration number of network training each time the DIP is used, so as to find a better solution. However, the strategy of re-initializing the network parameters greatly increases the uncertainty of network reconstruction, and the quality of reconstruction result is also easily affected.

To sum up, it can be seen that although DIP has good image denoising and image representation capabilities, it is easy to fall into local minima and the optimization process is time-consuming. On the other hand, CS reconstruction based on sparsity priors can find the optimal solution, but the reconstruction performance is unsatisfactory in severly ill-posed problems. However, it is surprising that the sparsity prior and DIP can compensate for each other. DIP can further optimize the images reconstructed by CS according to sparsity prior. Sparsity priors can be used to prevent the reconstruction results from falling into local minima. The bidirectional effect of DIP and sparsity prior will make the reconstruction result in the ill-posed problem reach an optimal solution. 

In this paper, we bridge the gap between sparsity prior and DIP, developing a fast alternating minimization algorithm based on the sparsity and deep image priors (Fama-SDIP). The proposed algorithm can avoid the reconstruction of HSI getting trapped in local minima.
Furthermore, there is no need to re-initialize the network parameters or increase iteration number of network training during the DIP training, which greatly improves the convergence speed and reconstruction performance.
According to the principle of CS, we constrain the images in their sparse domain for the purpose of accurate reconstruction. 
By using the split Bregman algorithm, we integrate the denoising characteristic of DIP into the optimization process of CS, and establish a fast alternating minimization algorithm, which can
achieve the purpose of fast reconstruction of the inverse problem.
To the best of our knowledge, this is the first time that the sparsity prior and DIP are explicitly utilized in the CASSI reconstruction problem. 

The remainder of this paper is organized as follows. Section II presents the CASSI forward model. Section III presents the principle of deep image prior. The reconstruction framework based on split Bregman algorithm is formulated in Section IV.
Section V presents the simulation and experimental results, and Section VI presents conclusion.

\section{Snapshot Measurement Model}
The CASSI system, which is mainly composed of a coded aperture, a prism and a gray-scale focal plane array, is used to acquire the 2D compressed measurements of the 3D spectral datacube. A concise schematic of the CASSI system is shown in Fig.~\ref{f1}. The spatial-spectral datacube of a scene is represented as $ \mathbf{X} \in \mathbb{R}^{M \times N \times L}  $, where $ M, N $ denote spatial dimensions, and $ L $ denotes spectral dimension, respectively. The encoded datacube is acquired when the spatial information of the scene is first modulated by a coded aperture $ \mathbf{T} \in \mathbb{R}^{M \times N} $. Then, the encoded datacube is shifted along the horizontal way after passing through a prism. Next, the detector measures the coded shifted spectral datacube, where the spectral information is integrated along the spectral dimension leading to the 2D compressive measurement $ \mathbf{Y} \in \mathbb{R}^{M \times (N + L -1)} $.

\begin{figure}[!t]
	\centering\includegraphics[width=8cm]{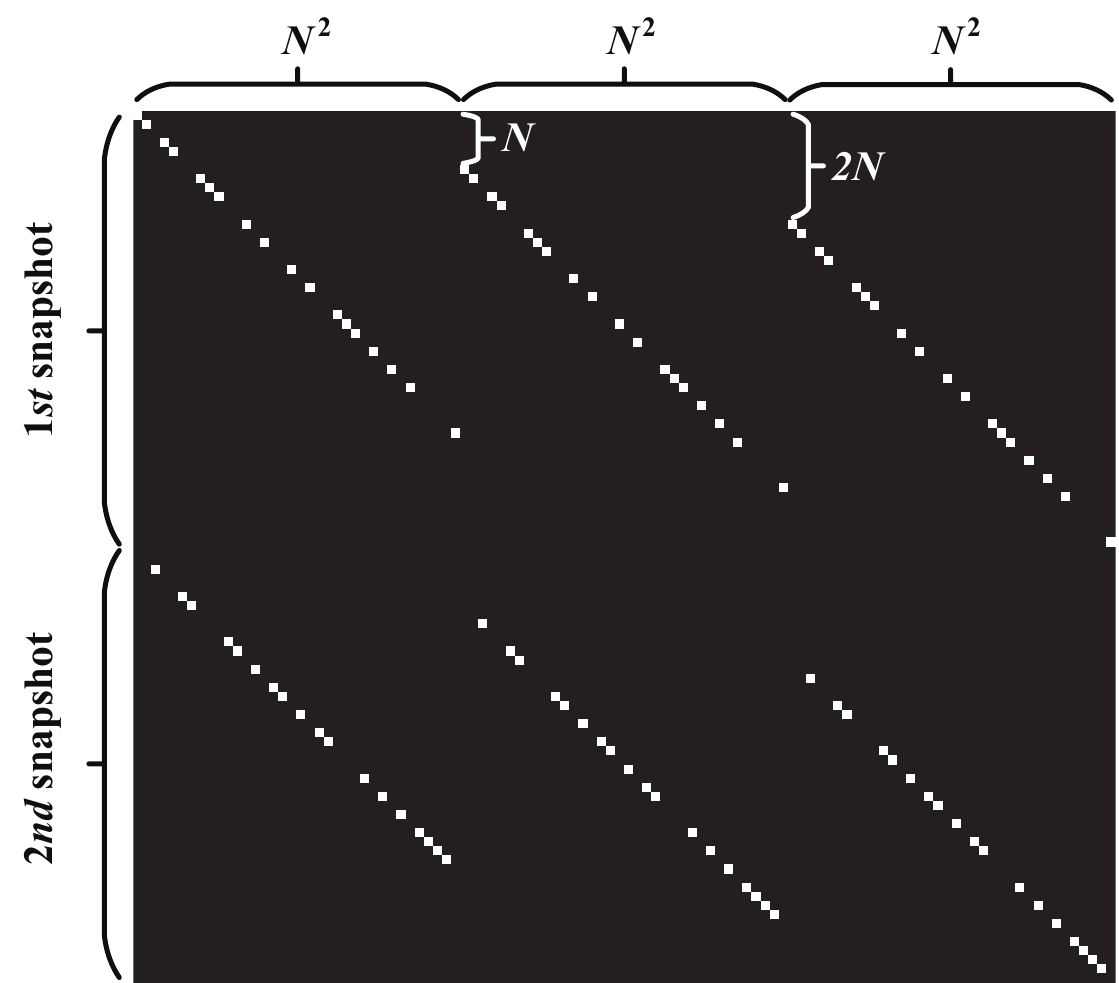} \caption{The system matrix $ \mathbf{H} $. The figure depicts the sensing of three spectral bands and two snapshots.}
	\label{f2}
\end{figure}

The vectorized representation of the datacube and snapshot measurement is represented as $ \bm{x} = \text{vec}(\mathbf{X}) \in \mathbb{R}^{MNL \times 1} $ and $ \bm{y} = \text{vec}(\mathbf{Y}) \in \mathbb{R}^{M (N + L -1) \times 1} $ respectively. Then, the forward model of CASSI can be written in the following matrix form:
\begin{equation}
\bm{y} = \mathbf{H}\bm{x} + \bm{\omega},
\end{equation}\label{e1}
where $ \mathbf{H} \in \mathbb{R}^{M (N + L -1) \times MNL} $ and $ \bm{\omega} \in \mathbb{R}^{M \times (N + L -1)} $ denote the sensing matrix and the sensing/system
noise, respectively. The system matrix denotes the joint effects of the coded aperture and the prism. In order to further explain the system matrix. We take a specific scenario with two snapshots as an example, and we set the datacube of dimensions $ M = N = 6, L = 3 $, and the transmittance of the coded apertures follows a Bernoulli distribution at 50\%. Then, the structure of this system matrix is shown in Fig.~\ref{f2}.  It can be observed that the matrix $ \mathbf{H} $ is sparse and highly structured, which provides robust conditions for CS reconstruction. 

The goal in CASSI is to reconstruct HSI from the compressive measurements by solving the ill-posed
inverse problem. 
Due to the sparsity property of HSI, it can be reconstructed according to the principle of CS. Suppose that the HSI can be sparsely represented on a orthonormal basis $ \mathbf{\Psi} $, i.e., $ \bm{x} = \mathbf{\Psi}\bm{\theta} $, where $ \bm{\theta} $ is composed of a small number of significant coefficients, referred to as the sparse coefficient vector of the HSI.  The sparse basis $ \mathbf{\Psi} $ is often generated by the Kronecker product of a two-dimensional-wavelet Symmlet-8 basis in the $ M - N $ plane and a 1D discrete Fourier transform basis along the $ L- $asis.
According to the CS theory, HSI can be reconstructed by solving the $ l_1 $ minimization problem:
\begin{equation}\label{3-1}
\arg \underset{\bm{x}}{\min}~\bigg\{\frac{1}{2}\|\bm{y}-\textbf{H}\bm{x} \|_2^2 + \xi_1\|\bm{\theta}\|_1  \bigg\},
\end{equation}
where $ \xi_1 $ is the regularization parameter, and sparse coefficient vector can be calculated as $ \bm{\theta} = \mathbf{\Psi}^T \bm{x} $.

\section{Deep Image Prior}
Deep image priors were originally proposed for image restoration tasks such as image denoising, image super-resolution, image inpainting\cite{dip1}. It shows that a randomly-initialized neural network can be used as a handcrafted prior, which can get excellent results in many image restoration tasks. Given a degraded image, a good image reconstruction can be obtained after DIP training. According to DIP, HSI can be estimated by a neural network, i.e., $ \bm{x}_{DIP}=f_\lambda(\bm{z}) $, where $ \bm{x}_{DIP} $ is the optimized result by DIP, and $ \bm{z} $ is a fixed random code vector with the same spatial size as $ \bm{x}_{DIP} $, and $ \lambda $ represents the parameters of the neural network. According to (\ref{3-1}), CASSI reconstruction problem can be solved by DIP:
\begin{equation}\label{3-2}
\arg \underset{\lambda}{\min}~\frac{1}{2}\|\bm{y}-\textbf{H}f_\lambda(\bm{z}) \|_2^2  .
\end{equation}

However, the problem of CASSI reconstruction is ill-posed, and if the network is trained without intervention, it will easy enter into a local minimum. In addition, increasing the number of training of the network can be helpful for reconstruction, however it will increase the computational burden and unavoidably lead to local minimum. To address this problem, we design an alternating iterative algorithm that incorporates both sparsity prior and DIP. 
Assuming that the output of a certain iteration is $ \bm{x} $, which can be used as the reference value of DIP.
Since the image is constrain in the sparse domain, the output of each iteration $ \bm{x} $ will gradually approach the optimal solution. Therefore, during DIP training, another fidelity term can be added, which is formulated as
\begin{equation}\label{3-3}
\arg \underset{\lambda}{\min}~\frac{1}{2}\|f_\lambda(\bm{z}) - \bm{x} \|_2^2  .
\end{equation}

The joint effect of (\ref{3-2}) and (\ref{3-3}) will enable DIP to denoise the output $ \bm{x} $. Due to the ``supervision'' of $ \bm{x} $, the network of DIP can get the best parameters without re-initializing, which greatly shortens the training time.

Combining (\ref{3-1}), (\ref{3-2}), (\ref{3-3}), the CASSI reconstruction problem can be expressed as
\begin{equation}\label{3-4}
\begin{aligned}
\arg \underset{\bm{x}, \lambda}{\min}~\bigg\{\frac{1}{2}\|\bm{y}-\textbf{H}\bm{x}\|_2^2  +\xi_1 \| \bm{\theta} \|_1 + \frac{1}{2} \|\bm{y}-\textbf{H} f_\lambda(\bm{z}) \|_2^2 &\\+ \frac{\eta}{2} \|f_\lambda(\bm{z})-\bm{x} \|_2^2 \bigg\},
\end{aligned}
\end{equation}
where $ \eta $ is the regularization parameter.

\section{Image reconstruction framework based on split Bregman algorithm}
\begin{figure*}[!t]
	\centering\includegraphics[width=16cm]{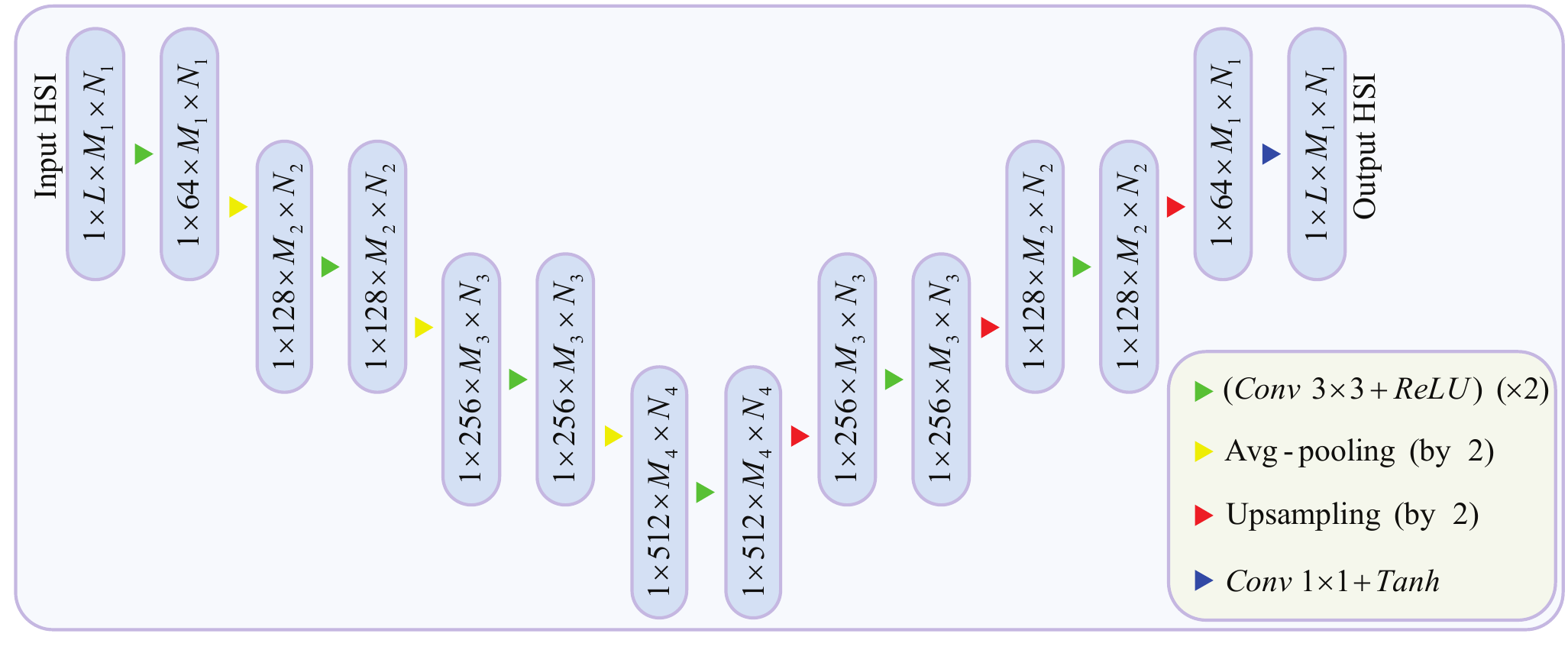} \caption{Schematic diagram of the neural network structure used for solving $ \textbf{Step 2} $. The input is progressively downsampled by factor of 2 at each scale  (e.g. $ M_4 = M_1 / 8 $).}
	\label{network}
\end{figure*}
In this section, we develop a fast alternating minimization algorithm based on split Bregman algorithm framework to solve for the HSI reconstruction problem in (\ref{3-4}), where the inverse problem is split into several sub-problems, and the sparse regularization term is replaced by the Bregman distance. Firstly, we introduce an auxiliary variable $ \bm{e} $ with the same dimension as $ \bm{y} $ to relax the $ l_2 $-norm by using the add-residual-back iterative scheme. To efficiently handle the non-differentiable norms, the arguments of the $ l_1 $-norm is replaced by the auxiliary variables $ \bm{c} = \bm{\theta} = \mathbf{\Psi}^T \bm{x} $. In addition, the add-residual-back iterative scheme is adopted to relax the $ l_1 $-norm into $ l_2 $-norm by introducing the auxiliary variable $ \bm{w} $, which indicates the difference between $ \bm{c} $ and $ \bm{\theta} $.  Similarly, we introduce an auxiliary variable $ \bm{b} $ with the same dimension as $ \bm{x} $ to relax the $ l_2 $-norm. Finally, the CASSI reconstruction problem is modified to
\begin{equation}\label{4-1}
\begin{aligned}
\hat{\bm{x}} &= \arg \underset{\bm{x},\lambda,\bm{e},\bm{c},\bm{w},\bm{b}}{\min}~\bigg\{\frac{1}{2}\|\bm{y}-\textbf{H}\bm{x}+\bm{e} \|_2^2   + \frac{\xi}{2} \|\bm{c} - \bm{\theta} - \bm{w} \|_2^2 \\&+\xi_1 \| \bm{c} \|_1 + \frac{1}{2} \|\bm{y}-\textbf{H} f_\lambda(\bm{z}) \|_2^2 + \frac{\eta}{2} \|f_\lambda(\bm{z})-\bm{x}+\bm{b} \|_2^2 \bigg\},
\end{aligned}
\end{equation}
where $ \xi $ is a regularization parameter.

The solution of (\ref{4-1}) includes the following four steps.

$ \textbf{Step 1} $. Update the vector $ \bm{x} $ and the auxiliary variable $ \bm{e} $:
\begin{equation}\label{4-2}
\begin{aligned}
&\bm{x}^{n+1} = \arg \underset{\Delta\bm{x}}{\min}~\bigg\{\frac{1}{2}\|\bm{y}-\textbf{H}(\bm{x}^n+\Delta\bm{x})+\bm{e}^n \|_2^2 \\&+ \frac{\xi}{2} \|\bm{c}^n - \bm{\theta}^n - \bm{w}^n \|_2^2 
+ \frac{\eta}{2} \|f_{\lambda^n}(\bm{z})-(\bm{x}^n+\Delta\bm{x})+\bm{b}^n \|_2^2 \bigg\},
\end{aligned}
\end{equation}
\begin{equation}\label{4-3}
\bm{e}^{n+1} = \bm{e}^n + \bm{y} - \textbf{H}\bm{x}^{n+1},
\end{equation}
where \textit{n} indicates the iteration number.

$ \textbf{Step 2} $. Update the parameters of neural network $ \lambda $ and the auxiliary variables $ \bm{b} $:
\begin{equation}\label{4-4}
\begin{aligned}
\lambda^{n+1} &= \arg \underset{\Delta\lambda}{\min}~\bigg\{\frac{1}{2} \|\bm{y}-\textbf{H} f_{(\lambda^n+\Delta\lambda)}(\bm{z}) \|_2^2 \\&+ \frac{\eta}{2} \|f_{(\lambda^n+\Delta\lambda)}(\bm{z})-\bm{x}^{n+1}+\bm{b}^{n} \|_2^2 \bigg\},
\end{aligned}
\end{equation}
\begin{equation}\label{4-5}
\bm{b}^{n+1} = \bm{b}^n + f_{\lambda^{n+1}}(\bm{z}) - \bm{x}^{n+1}.
\end{equation}

$ \textbf{Step 3} $. Update the auxiliary variables $ \bm{c} $ and $ \bm{w} $:
\begin{equation}\label{4-8-1}
\begin{aligned}
\bm{c}^{n+1} &= \arg \underset{\Delta\bm{c}}{\min}~\bigg\{\xi_1 \| (\bm{c}^n+\Delta\bm{c}) \|_1 \\&+ \frac{\xi}{2} \|(\bm{c}^n+\Delta\bm{c}) - \bm{\theta}^{n+1} - \bm{w}^n \|_2^2 \bigg\},
\end{aligned}
\end{equation}
\begin{equation}\label{4-9-1}
\bm{w}^{n+1} = \bm{w}^n + \bm{\theta}^{n+1} - \bm{c}^{n+1}.
\end{equation}

$ \textbf{Step 4} $. Return $\textbf{Step 1} $ until the algorithm converges or the maximum number of iterations is reached.

$ \textbf{Solve Step 1:} $ The quadratic optimization problem in (\ref{4-2})  has a closed-form solution formulated as:
\begin{equation}\label{4-7-1}
\begin{aligned}
\bm{x} &= (\textbf{H}^T\textbf{H} + \eta\textbf{I} + \xi\textbf{I})^{-1}[\textbf{H}^T(\bm{y}+\bm{e})+\eta(f_{\lambda}(\bm{z})+\bm{b})\\&+\xi\bm{\Psi}(\bm{c}-\bm{w})] .
\end{aligned}
\end{equation}
Due to fact that $ \textbf{H} $ is a fat matrix, the matrix inversion formula is employed to simplify the calculation by use of the Woodbury matrix identity:
\begin{equation}\label{4-7-2}
\begin{aligned}
(\textbf{H}^T\textbf{H} + \eta\textbf{I} + \xi\textbf{I})^{-1} &= (\eta + \xi)^{-1}\textbf{I}-(\eta + \xi)^{-1}\textbf{H}^T\\&(\textbf{I}+\textbf{H}(\eta + \xi)^{-1}\textbf{H}^T)^{-1}\textbf{H}(\eta + \xi)^{-1} .
\end{aligned}
\end{equation}

Plugging (\ref{4-7-2}) into (\ref{4-7-1}), the solution of $ \bm{x} $ can be obtained by
\begin{equation}\label{4-7-3}
\begin{split}
&\bm{\alpha} \overset{\rm{def}}{=} (\eta + \xi)^{-1}[\eta(f_{\lambda}(\bm{z})+\bm{b})+\xi\bm{\Psi}(\bm{c}-\bm{w})],
\\&\bm{x} = \bm{\alpha} + \textbf{H}^T(\bm{y}-\textbf{H}\bm{\alpha}+\bm{e}) \oslash (\rm{Diag}(\textbf{H}\textbf{H}^T)+\eta\textbf{I} + \xi\textbf{I}),
\end{split}
\end{equation}
where $ \oslash $ represents the operation of element-wise division, and $ \rm{Diag}() $ denotes the operation of extracting the diagonal elements.

$ \textbf{Solve Step 2:} $ For the implementation of DIP, we use a U-net \cite{U-net} without skip connections, which is a similar network structure as in \cite{PNP-DIP}. The schematic diagram of the neural network structure is shown in Fig.~\ref{network}.
The two quadratic optimization problems in (\ref{4-4}) can be equivalent to two loss functions of the neural network. The first loss function is to reduce the measurement error $ Loss_y = \left|\bm{y}-\textbf{H}f_\lambda(\bm{z}) \right|  $ according to the projection measurement value. The second loss function denotes as $ Loss_x = \left|f_\lambda(\bm{z}) - \bm{x} \right|  $, which can make the network output close to the reference value, so that the network jumps out of the local minimum. In addition, by restricting the output of the neural network to be close to $ \bm{x} $, the value of the auxiliary variable $ \bm{b} $ can also be minimized. Therefore, the loss function of the network is set as $ Loss = Loss_y + Loss_x $.
For simplicity, we do not add any balance weights between the two losses. In addition, $ \bm{z} $ generated by uniform noise is a fixed input variable of the neural network and of the same spatial size as $ \bm{x} $. Then, \textbf{Step 2} can be solved by training the neural network with back-propagation algorithm.

$ \textbf{Solve Step 3:} $
According to \cite{FIST,IST2}, the problem in (\ref{4-8-1}) can be solved by using the iterative soft thresholding algorithm, which can be formulated as
\begin{equation}\label{4-12}
\bm{c}^{n+1} = \text{Soft}\bigg\{\bm{c}^n - t\xi \times (\bm{c}^n-\bm{\theta}^{n+1}-\bm{w}^n), \frac{\xi_1}{\xi}\bigg\},
\end{equation}
where $ \xi_1 / \xi \geq 0 $, $ \bm{\theta}^{n+1} = \bm{\Psi}^T \bm{x}^{n+1} $, $ t $ is an appropriate stepsize, and $ \text{Soft}\{\cdot,\cdot\} $ is the soft-shrink operator. For an arbitrary vector $ \bm{v} \in \mathbb{R}^{N \times 1} $, the shrink operation is defined as 
\begin{equation}\label{4-13}
\text{Soft}\bigg\{\bm{v},\frac{\xi_1}{\xi}\bigg\} = \text{sgn}(\bm{v})\odot\max\bigg\{ \left|\bm{v}\right|-\frac{\xi_1}{\xi}, 0 \bigg\},
\end{equation}
where $ \text{sgn}() $ is Sign function.

For simplicity, (\ref{4-12}) can be rewritten as
\begin{equation}\label{4-12-2}
\bm{c}^{n+1} = \text{Soft}\bigg\{(1-t') \times \bm{c}^n +t' \times (\bm{\theta}^{n+1}+\bm{w}^n), \frac{\xi_1}{\xi}\bigg\},
\end{equation}
where $ t'=t\xi $. 

Following the abovementioned procedures, we have solved the above three steps, which can be solved efficiently by solving each sub-problem separately, leading to a stable solution. 

\begin{figure*}[!t]
	\centering\includegraphics[width=16cm]{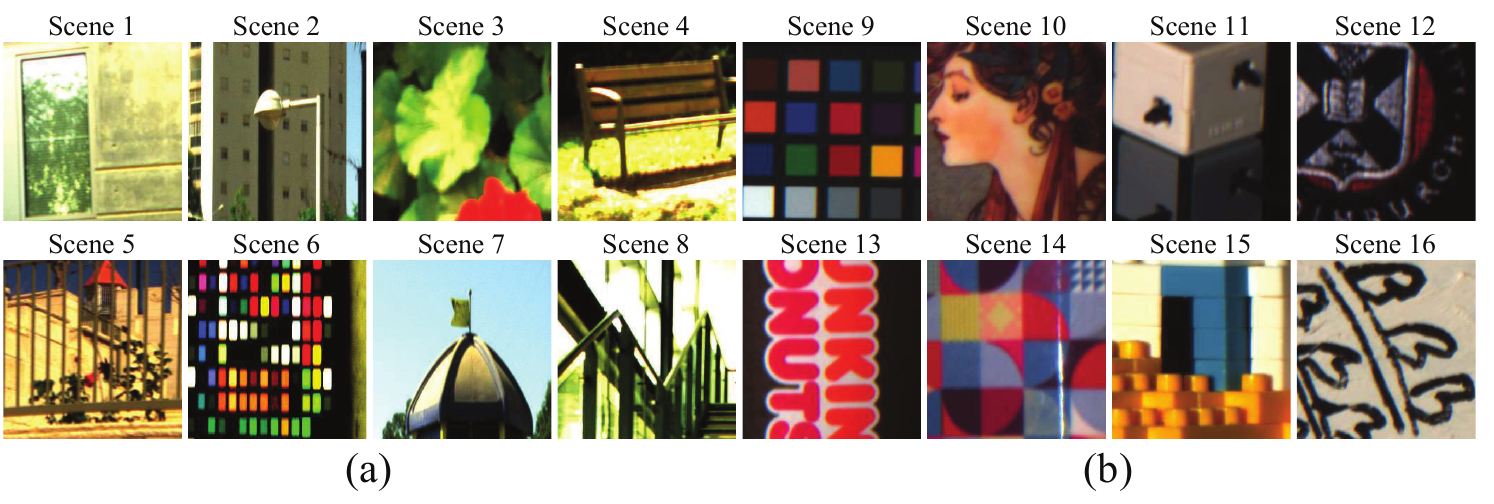} \caption{Spectral data scenes from (a) ICVL and (b) KAIST data sets used in simulations. }
	\label{f4-1}
\end{figure*}

\section{Results}
Extensive experiments are presented in this section to validate the performance of our proposed Fama-SDIP algorithm.  First, we conduct ample simulations to compare Fama-SDIP with other competitive methods including supervised-based methods. Then we built the CASSI experimental platform and perform real experiments on the platform to verify our proposed algorithm. In the simulations and in the real experiments, only one snapshot is used. In our simulations and real experiments, the parameters are uniformly set as follows: $ \xi_1=10 $, $ \xi=8 $, $ \eta=10 $, and $ t'=0.95 $. The initial value of $ \bm{x} $ is set as $ \bm{x}=\mathbf{H}^T\bm{y} $, and the initial value of $ f_\lambda(\bm{z}) $ is set as $ f_\lambda(\bm{z}) = \mathbf{H}^T\bm{y} $. In order to verify the robustness of the proposed algorithm, we use the same neural network model in both simulations and real experiments. It is worth noting that in our experiments, there is only one-pixel shift between the neighboring spectral channels, because for higher shift steps, the system will have more noise and require more accurate modeling\cite{hmodel}.
In the training process of DIP, Adam\cite{Adam} optimizer $ (\beta_1=0.9$ and $\beta_2=0.999) $ is used. The learning rate is set to be 0.002, and the weight decay is set as $ 1\text{e}^{-4} $.  In our implementation, the maximum iteration number of Fama-SDIP is set as 45, and the DIP inner loop is set as 100 times in each iteration.
In order to ensure the robustness of the algorithm, we should ensure that the initial $ Loss_y $ is close to the end $ Loss_y $ of the previous DIP output each time solve Step 2, otherwise we first optimize the network to make it close to the end $ Loss_y $ of the previous DIP output.

\subsection{Simulation Results}
\begin{figure*}[!t]
	\centering\includegraphics[width=18cm]{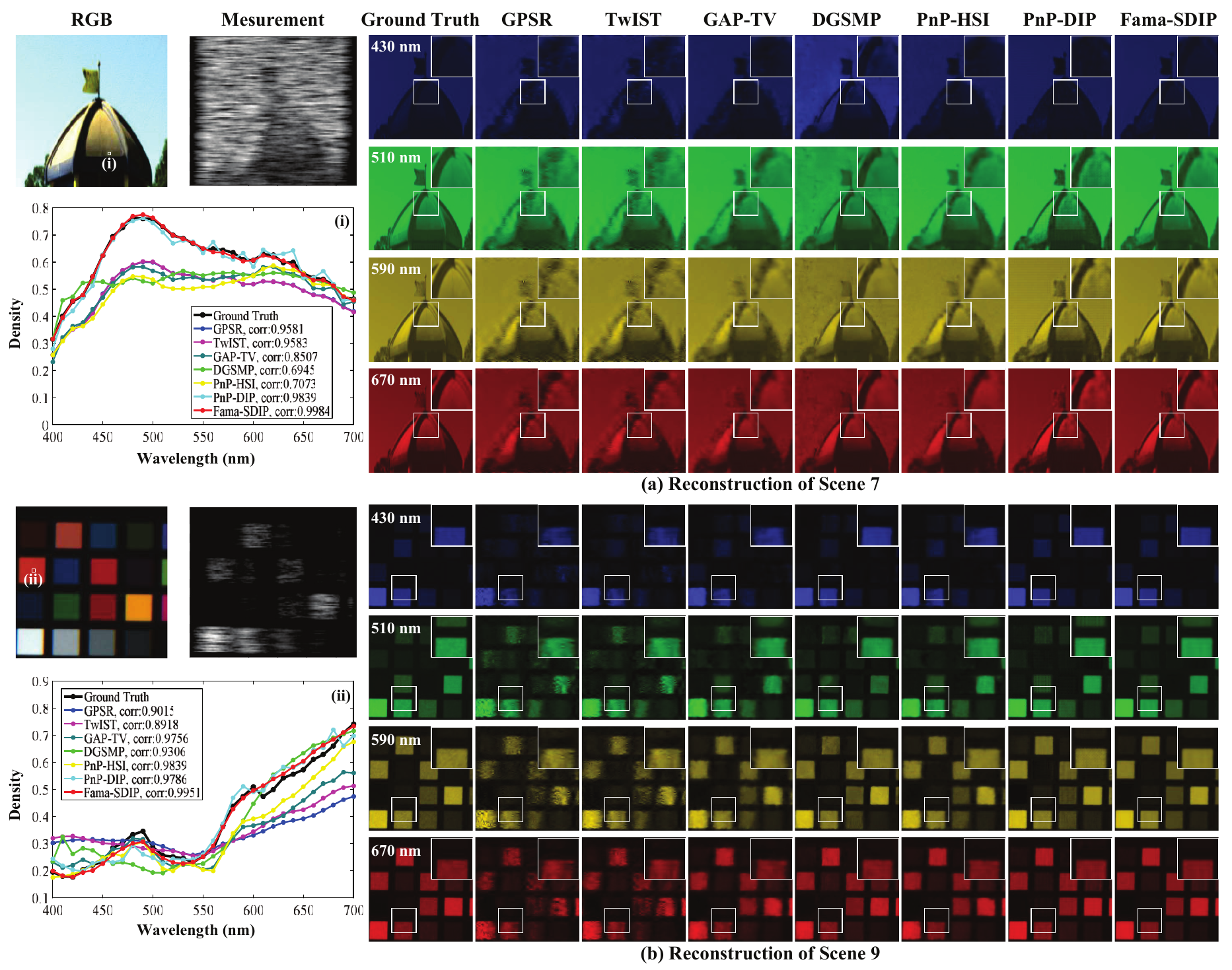} \caption{Reconstructed simulation HSIs comparisons of Scene 7 and 9 with 4 out of 31 spectral channels.
		The reconstructed spectral curves on selected regions are shown for comparing the spectral accuracy of different algorithms.  The correlation of the reconstructed spectra is shown in the legends.}
	\label{f4-2}
\end{figure*}

The publicly available datasets ICVL\cite{ICVL} and KAIST\cite{KAIST} are employed for simulations. The ICVL data with spatial size $ 1392 \times 1300 $ and the KAIST data with spatial size $ 2704 \times 3376 $ have same 31 spectral bands, and both of their wavelengths are in the range from 400 to 700 nm at a step of 10 nm.
In each data set, we select eight scenes identical to those in \cite{PnP}. Then, we cropped the data set to spatial sizes of $ 256 \times 256 $ as shown in Fig~\ref{f4-1}. 
We compare our proposed algorithm with other leading algorithms, including three traditional algorithms, i.e. GPSR\cite{GPSR}, TwIST\cite{TwIST}, GAP-TV\cite{GAP-TV}, two DL methods based on supervised learning, i.e. PnP-HSI\cite{PnP}, DGSMP\cite{DGSMP}, one DL methods based on unsupervised learning, i.e. PnP-DIP\cite{PNP-DIP}.
For the sake of fair comparison, we divide each data by its maximum value so that its image peak is 1, and we use the same real coded aperture as in\cite{DGSMP,PNP-DIP} to generate the compressive measurements.
We apply the peak signal to noise ratio (PSNR) and structural similarity (SSIM)\cite{ssim} as the objective quality metrics to evaluate the quality of reconstructed spectral datacube.
\begin{table*}[!t]
	\caption{Quantitative results on 16 Simulation Scenes (8 from ICVL and 8 from KAIST). PSNR and SSIM are reported. \label{table1}}
	\centering
	\begin{tabular}{|c||c||c||c||c||c||c||c|}
		\hline
		Algorithms & GPSR & TwIST & GAP-TV & DGSMP & PnP-HSI & PnP-DIP & Proposed Fama-SDIP \\
		\hline
		Scene 1 & 27.16, 0.879 & 27.09, 0.879 & 28.95, 0.913 & 23.94, 0.825 & 29.36, 0.909 & 30.19, 0.915 & 34.74, 0.965 \\
		\hline
		Scene 2 & 23.04, 0.848 & 23.39, 0.851 & 25.50, 0.893 & 27.71, 0.923 & 26.96, 0.911 & 32.41, 0.948 & 36.13, 0.979 \\
		\hline
		Scene 3 & 25.79, 0.937 & 26.43, 0.941 & 38.67, 0.988 & 34.66, 0.962 & 38.83, 0.986 & 38.54, 0.981 & 43.36, 0.994 \\
		\hline
		Scene 4 & 26.20, 0.871 & 27.00, 0.880 & 29.28, 0.920 & 29.76, 0.924 & 29.87, 0.923 & 31.24, 0.929 & 33.99, 0.959 \\
		\hline
		Scene 5 & 21.14, 0.709 & 21.30, 0.716 & 22.71, 0.778 & 25.46, 0.872 & 23.44, 0.796 & 28.35, 0.903 & 30.48, 0.939 \\
		\hline
		Scene 6 & 21.55, 0.732 & 21.58, 0.732 & 23.94, 0.831 & 24.88, 0.858 & 24.78, 0.847 & 28.20, 0.894 & 32.48, 0.958 \\
		\hline
		Scene 7 & 26.32, 0.902 & 26.28, 0.901 & 28.72, 0.938 & 23.87, 0.777 & 29.91, 0.944 & 31.24, 0.929 & 35.03, 0.975\\
		\hline
		Scene 8 & 28.56, 0.895 & 29.51, 0.902 & 31.28, 0.930 & 30.38, 0.938 & 32.04, 0.937 & 34.87, 0.962 & 37.12, 0.978 \\
		\hline
		Scene 9 & 20.78, 0.766 & 22.41, 0.797 & 26.35, 0.907 & 27.29, 0.888 & 28.81, 0.939 & 30.65, 0.905 & 36.38, 0.977\\
		\hline
		Scene 10 & 24.66, 0.852 & 24.75, 0.854 & 28.09, 0.922 & 20.37, 0.653 & 28.03, 0.904 & 28.90, 0.911 & 32.45, 0.960\\
		\hline
		Scene 11 & 26.86, 0.887 & 27.05, 0.889 & 27.59, 0.926 & 30.44, 0.939 & 30.07, 0.951 & 32.37, 0.927 & 37.33, 0.980\\
		\hline
		Scene 12 & 21.70, 0.825 & 21.20, 0.706 & 23.63, 0.809 & 25.29, 0.836 & 24.39, 0.832 & 30.34, 0.932 & 32.95, 0.970\\
		\hline
		Scene 13 & 18.66, 0.711 & 19.45, 0.735 & 23.26, 0.857 & 23.44, 0.857 & 24.82, 0.881 & 30.82, 0.923 & 34.19, 0.974\\
		\hline
		Scene 14 & 24.17, 0.866 & 24.87, 0.877 & 27.36, 0.932 & 22.63, 0.760 & 28.02, 0.939 & 29.05, 0.926 & 32.13, 0.973\\
		\hline
		Scene 15 & 22.97, 0.805 & 23.30, 0.812 & 26.22, 0.904 & 25.73, 0.832 & 26.74, 0.918 & 29.75, 0.902 & 33.73, 0.972\\
		\hline
		Scene 16 & 19.03, 0.715 & 19.30, 0.725 & 19.16, 0.747 & 24.04, 0.839 & 20.59, 0.794 & 28.88, 0.921 & 32.38, 0.963\\
		\hline
		Average & 23.66, 0.825 & 24.06, 0.825 & 26.92, 0.887 & 26.23, 0.855 & 27.92, 0.901 & 30.99, 0.926 & 34.68, 0.970\\
		\hline
	\end{tabular}
\end{table*}
The performance comparisons on the sixteen benchmark scenes are given in Table \ref{table1}, using different algorithms, i.e., GPSR\cite{GPSR}, TwIST\cite{TwIST}, GAP-TV\cite{GAP-TV}, DGSMP\cite{DGSMP}, PnP-HSI\cite{PnP}, PnP-DIP\cite{PNP-DIP} and our proposed Fama-SDIP.
It can be seen that the PSNR and SSIM values of our proposed Fama-SDIP are much higher than other reconstruction algorithms.
Since there is just one-pixel shift between the neighboring spectral channels, the compression ratio is increased compared with the case of two-pixels shift, so the supervised learning method DGSMP has a significant decrease in the reconstruction accuracy. Compared with the unsupervised learning method PnP-DIP, the proposed Fama-SDIP shows an improvement of up to 3.6dB in average PSNR for reconstructions obtained.
Figure \ref{f4-2} plots selected reconstructed scenes of Fama-SDIP compared with GPSR, TwIST, GAP-TV, DGSMP, PnP-HSI and PnP-DIP. 
We can observe from the reconstructed HSIs and the magnified patches within the white boxes that previous methods are less favorable for recovering HSI details, and their reconstruction results all produce different degrees of image blur and artifacts. In contrast, Fama-SDIP produces sharper borders and better image details because the image is optimized towards the optimal solution under the bidirectional constraints of DIP and sparsity prior, leading to the state-of-the-art results on both PSNR and SSIM.
Furthermore, we also plot the reconstructed spectral curves of
two selected regions and calculate the correlations with the reference spectra. It can be seen that Fama-SDIP provides more accurate spectra.

\subsection{Running Time}
Table \ref{table2} compares the running times time on sixteen data of training the models and reconstructing the HSIs by the proposed Fama-SDIP method and other methods. All of the simulations are carried out on a computer with Intel
Core i7-8700K CPU, 16GB of RAM, and an Nvidia RTX 2080Ti GPU.
Although the proposed method dose not improve in the reconstruction time compared with the supervised learning method represented by DGSMP, it takes more than 10 days for DGSMP to train a model, which will be a great challenge for practical use. Compared to the unsupervised learning method represented by PnP-DIP, the proposed Fama-SDIP method can achieve more than 10-fold speedup and provides better results.
\begin{table}[!t]
	\caption{Runtimes of training or reconstruction. \label{table2}}
	\centering
	\begin{tabular}{|c||c||c|}
		\hline
		Algorithms & Training & Reconstruction \\
		\hline
		GPSR & - & 29.07 min \\
		\hline
		TwIST & - & 31.79 min \\
		\hline
		GAP-TV & - & 2.32 min \\
		\hline
		DGSMP & 10 days & 0.13 sec \\
		\hline
		PnP-HSI & 3 days & 2.23min \\
		\hline
		PnP-DIP & - & 115.63 min \\
		\hline
		Fama-SDIP & - & 10.75 min \\
		\hline
	\end{tabular}
\end{table}

\subsection{Real data Results}
\begin{figure*}[!t]
	\centering\includegraphics[width=15cm]{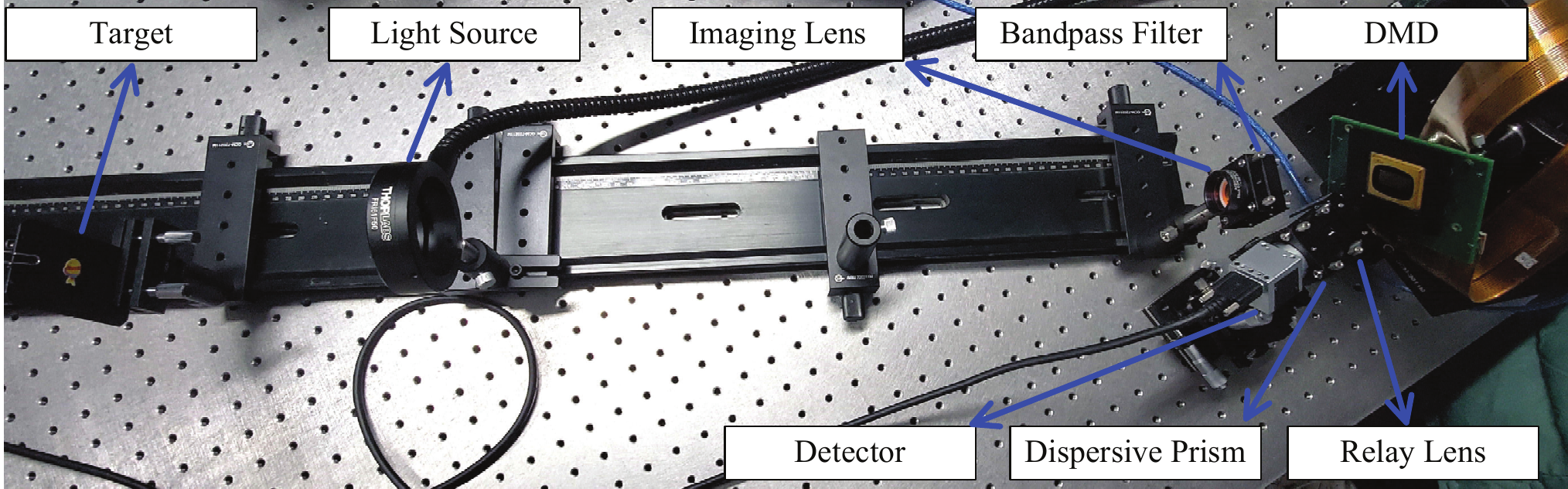} \caption{Our proof-of-concept test bed of CASSI.}
	\label{f5-1}
\end{figure*}

In this section, we apply the proposed Fama-SDIP algorithm into our real CASSI systems as shown in Fig.~\ref{f5-1}. The system includes a light source (Zolix GLORIA-X500A), an imaging lens (Thorlabs AC254-100-A-ML), bandpass filters (Daheng Optics GCC-300117 \& GCC-211002), a digital micromirror device (DMD) (Texas Instruments DLP9500), a relay lens (Edmund Optics \#45-762), a dispersive prism (double Amici prism designed in \cite{prism}), and a detector (Basler acA2040-90μm).
Note that the parameters we used in the real experiment are exactly the same as in the simulation.
In order to obtain robust reconstruction results, we can firstly use sparsity prior in the image reconstruction framework to obtain a result, which serves as a warm starting point for Fama-SDIP.

\begin{figure*}[!t]
	\centering\includegraphics[width=18cm]{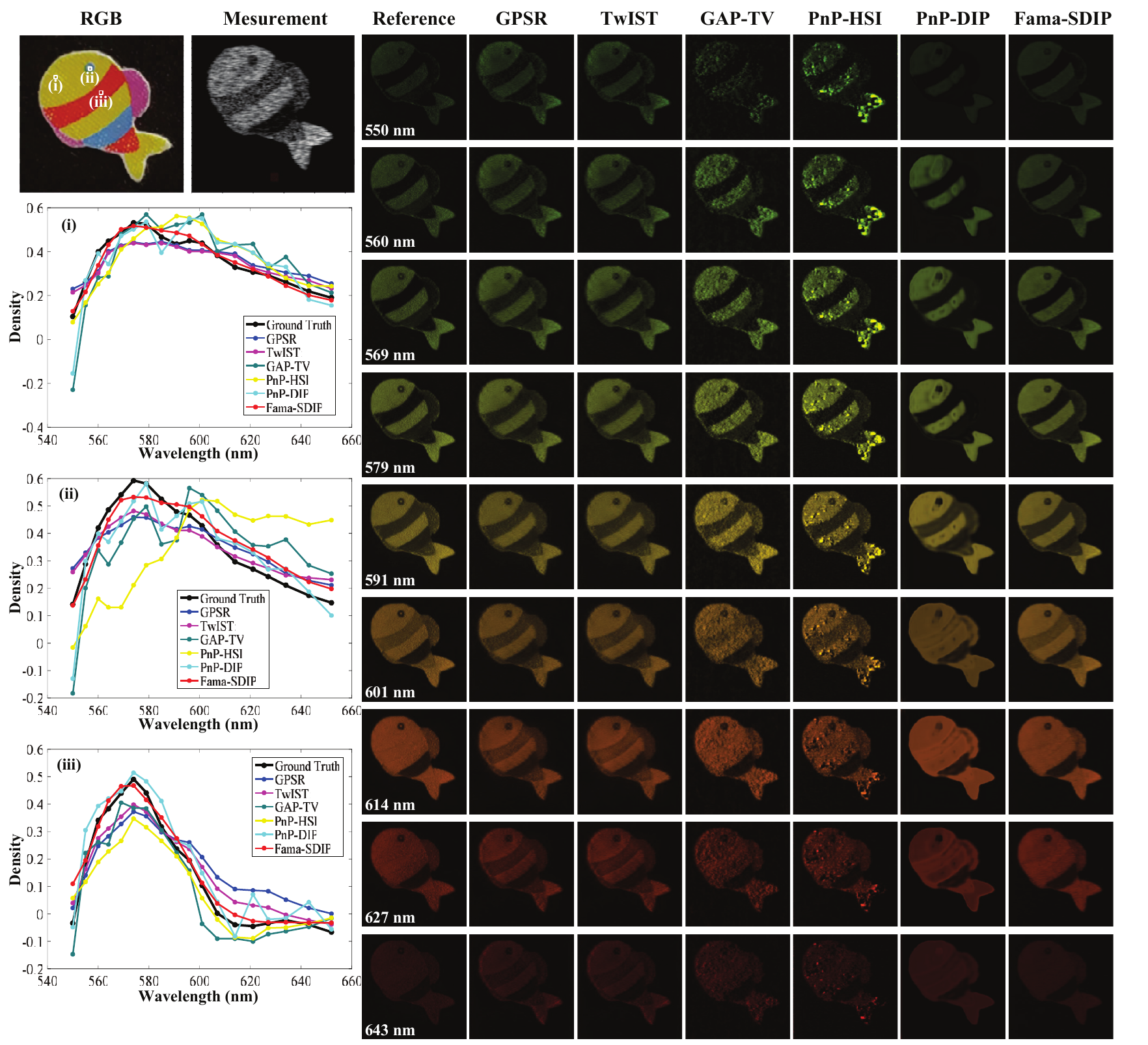} \caption{Reconstructed spectral images of real data.}
	\label{f5-2}
\end{figure*}

The spectral cube in our experiment contains 18 spectral channels with the spatial size of 512$ \times $512, and its wavelengths are in the range from 550nm to 652nm. 
Since the data set in real scenes is difficult to obtain, the DGSMP algorithm is hard to implement. In addition, PnP-HSI algorithm is using the pretrained HSI denoising network on the simulation data, so it can be used in our comparative experiments. As shown in Fig~\ref{f5-2}, the reconstruction results of different algorithms are displayed. It can be seen that compared with other five algorithms, our reconstruction has less artifacts and less noise. Furthermore, our reconstruction is able to distinguish the spectral features of different channels, which are clearly displayed around the wavelength of 614nm. This also means that our proposed algorithm can better handle the ill-posed inverse reconstruction problem. It is worth noting that the reconstructed images of PnP-HSI are worse than that attained by GAP-TV, which shows that the algorithm based on supervised learning may not work in real scenes.
In addition, we select three regions to display the corresponding recovered spectral curves, and the results also show that our proposed algorithm has higher spectral accuracy.

\section{Conclusion}
This paper developed a fast alternating minimization algorithm for coded aperture snapshot spectral imaging. Through synergistically utilizing the sparsity and deep image priors, the ill-posed reconstruction problem can be solved by using split Bregman algorithm. The proposed method can effectively reconstruct HSI within a relatively short period of time and does not need any training dataset. We verified the effectiveness and robustness of the proposed algorithm in both of simulations and real experiments, in which we use the same set of parameters to achieve state-of-art results.

\section*{Acknowledgments}
The author would like to thank the China Scholarship Council (202106030517).


\end{document}